# Universal dynamical decoupling of a single solid-state spin from a spin bath

G. de Lange[1], Z.H. Wang[2], D. Ristè[1], V.V. Dobrovitski[2], and R. Hanson[1*]

1. *Kavli Institute of Nanoscience Delft, Delft University of Technology,*

   *P.O. Box 5046, 2600 GA Delft, The Netherlands*

2. *Ames Laboratory and Iowa State University, Ames, Iowa 50011, USA*

Abstract

Controlling the interaction of a single quantum system with its environment is a fundamental challenge in quantum science and technology. We dramatically suppress the coupling of a single spin in diamond with the surrounding spin bath by using double-axis dynamical decoupling. The coherence is preserved for arbitrary quantum states, as verified by quantum process tomography. The resulting coherence time enhancement is found to follow a general scaling with the number of decoupling pulses. No limit is observed for the decoupling action up to 136 pulses, for which the coherence time is enhanced more than 25 times compared to spin echo. These results uncover a new regime for experimental quantum science and allow to overcome a major hurdle for implementing quantum information protocols.

In the last decade, manipulation and measurement of single quantum systems in the solid state has been achieved *(1, 2)*. This control has promising applications in quantum information processing *(3, 4)*, quantum communication *(5)*, metrology *(6)*, and ultra-sensitive magnetometry *(7, 8)*. However, uncontrolled interactions with the surroundings inevitably lead to decoherence of the quantum states *(9)* and pose a major hurdle for realizing these technologies. Therefore, the key challenge in current experimental quantum science is to protect individual quantum states from decoherence by their solid-state environment.

If a quantum system can be controlled with high fidelity, dynamical decoupling can be exploited to efficiently mitigate the interactions with the environment *(10, 11, 12)*. By reversing the evolution of the quantum system at specific times using control pulses, the effect of the environment accumulated before the pulse is cancelled during the evolution after the pulse. When viewed at the end of the control cycle, the quantum system will appear as an isolated system that is decoupled from its environment. Thanks to very recent progress in quantum control speed and precision *(13, 14)*, we are now able to unlock the full power of dynamical decoupling at the level of a single spin.

We focus on electron spins of single nitrogen-vacancy (NV) defect centers in diamond coupled to a spin bath (Fig. 1A). NV center spins can be optically imaged, initialized and read out, as well as coherently controlled at room temperature (Fig. 1B). These favorable properties have been exploited in the past years to gain deeper insight into spin decoherence *(15, 16)* as well as for demonstrating basic quantum information protocols at room temperature *(17, 18)*.

We use nanosecond microwave pulses to manipulate single NV spins. In order to raise the fidelity of our control to the required level for efficient decoupling, we have fabricated on-chip coplanar waveguide (CPW) transmission lines using electron beam lithography (Fig. 1A). The high bandwidth of the CPW *(13)* combined with efficient suppression of reflections and fine-tuned pulse calibration *(14)* allows fast and precise manipulation of the NV spin (Fig. 1B), leading to process fidelities of 99% for the basic control pulses needed for dynamical decoupling *(14)*.

The coherent dynamics of an NV spin are strongly influenced by the coupling to neighboring spins (the spin bath) *(15, 16)*. Because such spin environments are very common in the solid state, our results are directly relevant for other solid-state quantum bits such as spins in quantum dots *(19, 20)* and donors in silicon *(4, 21)*. For the NV centers studied here, the bath is comprised of electron spins localized on nitrogen impurity atoms. Resonant interactions (flip-flops) between the bath spins and the NV spin are suppressed due to a large energy mismatch *(16)*. Therefore, the impact of the spin bath on the NV spin is limited to dephasing and can be described as a random magnetic field $B(t)$ that is directed along the NV's quantization axis. The value of $B(t)$ is determined by the state of the environment. We model the bath field $B(t)$ by an Ornstein-Uhlenbeck process with the correlation function $C(t) = <B(0)B(t)> = b^2 \exp(-|t|/\tau_C)$, where $b$ is the coupling strength of the bath to the spin and $\tau_C$ is the correlation time of the bath which measures the rate of flip-flops between the bath spins due to the intra-bath dipolar coupling *(14, 22)*.

The values of the parameters describing the bath field are extracted from experiments. The bath-induced dephasing during free evolution has a Gaussian envelope $S(t)=\exp(-b^2t^2/2)$, which yields the value for $b$ *(14)*; we find $b = (3.6 \pm 0.1)$ µs$^{-1}$ for NV1 (Fig. 1C), and $b = (2.6 \pm 0.1)$ µs$^{-1}$ for NV2 *(14)*. The quasi-static dephasing can be undone using a spin echo (SE) technique (Figure 2A), revealing the much slower decay of spin coherence caused by the dynamics of the spin bath. The spin echo signal decays as $SE(t) = \exp[-(t/T_2)^3]$, characteristic for a slowly fluctuating spin bath with $\tau_C = T_2^3 b^2/12 \gg 1/b$ *(22)*. The values we find for $\tau_C$, $(25\pm3)$ µs for NV1 ($T_2 = (2.8 \pm 0.1)$ µs) and $(23\pm3)$ µs for NV2 ($T_2 = (3.5 \pm 0.2)$ µs), confirm this. The spin echo decay time $T_2$ is often considered as the coherence or memory time of the system. We take $T_2$ as the starting point and demonstrate that the coherence time can be dramatically prolonged by dynamically decoupling the spin from the surrounding spin bath.

We first explore the potential of dynamical decoupling by extending the spin-echo (SE) pulse sequence to periodic repetitions of the Carr-Purcell-Meiboom-Gill (CPMG) cycle (Fig. 2A). The decoupling performance is characterized by measuring the state fidelity $F_s = \langle \psi_i | \rho_m | \psi_i \rangle$, where $|\psi_i\rangle$ is the expected (ideal) state after applying the sequence and $\rho_m$ the measured density matrix of the actual state. While the coherence has vanished after 4 microseconds for the SE case, we observe that the 8-pulse CPMG sequence preserves the coherence almost completely during this same time.

The optimal decoupling sequence for a quantum system depends on the coupling to its environment and the dynamics within the environment itself. In Ref. *(23)*, non-periodic inter-pulse spacing, now called the UDD sequence, was found to achieve a strong improvement in decoupling efficiency over periodic pulse spacing in the case of environmental noise spectra with a hard cut-off; this was experimentally verified in Refs. *(24,25)*. Recent theory *(26, 27)*, however, suggests that periodic, CPMG-like pulse spacing is ideal for decoupling from an environment with a soft cut-off. We investigate the efficiency of these different protocols in decoupling a single spin from a spin bath environment (Fig. 2B) and observe that CPMG outperforms UDD for all numbers of pulses investigated in both simulations and experiments (Fig. 2B, right panel). These findings are in agreement with our model of a Lorentzian bath noise spectrum, which exhibits a soft cut-off *(14)*.

For applications in quantum information processing, it is essential that the decoupling protocol is universal, meaning it can preserve coherence for arbitrary quantum states. As pulse errors can severely degrade the coherence, universal decoupling requires robustness to pulse errors for all possible quantum states. In contrast, protocols that employ single-axis decoupling such as CPMG optimally preserve only a limited range of quantum states, while for other quantum states the pulse errors accumulate rapidly with increasing number of control pulses. In Fig. 2C we demonstrate this experimentally by comparing the decay curves of superposition states aligned ($|x\rangle$) and perpendicular ($|y\rangle$) to the CPMG decoupling axis. Even though the fidelity of the single-pulse control is very high *(14)* the remaining small errors cause a significant loss of decoupling fidelity for state $|y\rangle$ when the number of operations is increased to 12 pulses; this effect is accurately reproduced by simulations (Fig. 2C) *(14)*.

The use of sequences containing decoupling pulses over two axes, such as XY4 (Fig. 2D) *(28)* avoids this selective robustness to pulse errors and can compensate certain systematic pulse errors and coherent resonant perturbations without increasing control overhead. We find that XY4 is indeed capable of preserving both quantum states $|x\rangle$ and $|y\rangle$ (Fig. 2D).

We study the decoupling performance in more detail with the use of quantum process tomography (QPT), which allows for a complete characterization of any quantum process *(29)*. Figure 3 shows the experimental QPT results for XY4 with $N = 8$ operations, at different free evolution times. For a free evolution time of 4.4 microseconds, much longer than $T_2$, the measured process matrix $\chi$ is in excellent agreement with the ideal process of identity that is expected for perfect universal decoupling.

By taking snapshots of the process for different free evolution times, we can monitor how decoherence affects the quantum states. We observe that after $t = 10$ μs the process element corresponding to identity has decreased, while the $\sigma_z$-$\sigma_z$ element grows. After 20 μs these elements have approximately equal amplitudes. This behavior is characteristic for pure, off-diagonal dephasing *(29)* and is consistent with our model of the environment, in which the magnetic dipolar coupling with the bath leads to phase randomization. The independently measured energy relaxation time $T_1 > 1$ ms *(14)* confirms that longitudinal decay is not relevant in this regime.

Finally, we investigate how the coherence time scales with the number of control pulses. A detailed theoretical analysis shows that for $N$ perfect pulses, the decoupling fidelity decays as $F(t) = \exp[-A\, N\, t^3/(2N\tau_C)^3]$, where the total free evolution time $t = 2N\tau$ and $2\tau$ is the inter-pulse distance *(14)*. For the XY4 sequence, we find $A=(2/3)\, b^2\tau_C^2$ for both large and small $N$. The theory predicts two interesting features: first, the decay follows the universal form $\exp[-(t/T_{coh})^3]$ for all $N$, and second, the $1/e$ decay time scales as $T_{coh}(N) = T_2 N^{2/3}$.

In Fig. 4A we show XY4 decoupling for $N = 4, 16, 72$, as well as the spin echo for comparison. These data indicate that the $1/e$ decay time indeed scales with the number of pulses. For a thorough comparison with the theory we renormalize the time axis to $T_2 N^{2/3}$ (Fig. 4B). We find that all data collapse onto a single curve in line with the prediction. Then, we plot the $1/e$ decay time of coherence of NV1 and NV2, and fit to the expected scaling law. The data of both NV centers show excellent agreement with the theory over a range in $N$ spanning two orders of magnitude. For the longest sequence applied (136 pulses) the coherence time is increased by a factor of 26.

An interesting question is whether there is a limit to the coherence enhancement that can be achieved with dynamical decoupling. Our results demonstrate that we can prolong the spin coherence beyond the bath correlation time $\tau_C$. Also, the nuclear spin bath, which would affect the NV dynamics on a 5-microsecond timescale for the magnetic field used here *(15)*, is efficiently decoupled from the NV spin. In fact, the theory indicates no fundamental limit to the coherence time. In practice the decoupling efficiency will be limited by the minimum inter-pulse delay (of the order of the pulse widths), and the longitudinal relaxation time.

Since the spin bath environment is common to solid-state quantum bits, our findings can be transferred to other promising systems such as spins in quantum dots *(3, 19, 20)* and donors in silicon *(4, 21)*. Furthermore, the performance of spin-based magnetometers can greatly benefit from this work, since the magnetic field sensitivity scales with the coherence time *(7, 8)*. Finally, dynamical decoupling can be applied to protect entangled states, which are at the heart of quantum information science.

* To whom correspondence should be addressed. E-mail: r.hanson@tudelft.nl.

References and Notes

1. J. Clarke, F. K. Wilhelm, *Nature* **453,** 1031-1042 (2008).
2. R. Hanson, D. D. Awschalom, *Nature* **453**, 1043-1049 (2008).
3. D. Loss, D. P. DiVincenzo, *Phys. Rev. A* **57,** 120 (1998).
4. B. E. Kane, *Nature* **393,** 133-137 (1998).
5. L. Childress, J. M. Taylor, A. S. Sørensen, M. D. Lukin, *Phys. Rev. Lett.* **96**, 070504 (2006).
6. J. A. Jones *et al*., *Science* **324**, 1166-1168 (2009).
7. C. Degen, *Appl. Phys. Lett.* **92**, 24311-24319 (2008).
8. J. M. Taylor *et al*., *Nature Physics* **4**, 810-816 (2008).
9. W. H. Zurek, *Nature Physics* **5**, 181-188 (2009).
10. L. Viola, L. Knill, S. Lloyd, *Phys. Rev. Lett*. **82,** 2417-2421 (1999).
11. D. Vitali, P. Tombesi, *Phys. Rev. A* **59,** 4178-4186 (1999).
12. K. Khodjasteh, D. Lidar, *Phys. Rev. Lett*. **95**, 180501 (2005).
13. G. D. Fuchs, V. V. Dobrovitski, D. M. Toyli, F. J. Heremans, D. D. Awschalom, *Science* **326,** 1520-1522 (2009).
14. See supporting material.
15. L. Childress *et al*., *Science* **314**, 281-285 (2006).

16. R. Hanson, V. V. Dobrovitski, A. E. Feiguin, O. Gywat. D. D. Awschalom, *Science* **320**, 352-355 (2008).

17. L. Jiang *et al.*, *Science* **326**, 267-272 (2009).

18. P. Neumann *et al.*, *Nature Physics* **6**, 249-253 (2010).

19. R. Hanson, L. P. Kouwenhoven, J. R. Petta, S. Tarucha, L. M. K. Vandersypen, *Rev. Mod. Phys.* **79,** 1217-1265 (2007).

20. H. Bluhm *et al.*, *arXiv:1005.2995v1* (2010).

21. J. J. L. Morton *et al.*, *Nature* **455**, 1085-1088 (2008).

22. J. Klauder, P. Anderson, *Phys. Rev.* **125,** 912-932 (1962).

23. G. Uhrig, *Phys. Rev. Lett*. **98**, 100504 (2007).

24. M. J. Biercuk *et al.*, *Nature* **458**, 996-1000 (2009).

25. J. Du *et al.*, *Nature* **461**, 1265-1268 (2009).

26. S. Pasini, G.S. Uhrig, *Phys. Rev. A* **81**, 012309 (2010).

27. L. Cywinski, R. M. Lutchyn, C. P. Nave, and S. Das Sarma, *Phys. Rev. B* **77**, 174509 (2008).

28. T. Gullion, D. Baker, M. S. Conradi, *J. Mag. Res.* **89**, 479-484 (1990).

29. M. A. Nielsen, I.L. Chuang, *Quantum Computation and Quantum Information*. (Cambridge University Press, Cambridge, 2000).

30. We acknowledge support from the Defense Advanced Research Projects Agency, the Dutch Organization for Fundamental Research on Matter (FOM) and the Netherlands Organization for Scientific Research (NWO). Work at the Ames Laboratory was supported by the U.S. Department of Energy Basic Energy Sciences under contract no. DE-AC02-07CH11358.

Figure Captions

**Fig. 1. Quantum control of a single spin in diamond.**

(**A**) Left: A Nitrogen-Vacancy defect is formed by a single substitutional nitrogen ($^{14}$N) atom and an adjacent vacancy (V). The NV electron spin (orange arrow) is coupled to the host $^{14}$N nuclear spin (blue arrow) through the hyperfine interaction. Middle: The NV center is surrounded by a bath of electron spins located at sites of substitutional nitrogen atoms in the diamond lattice *(16)*. Right: Confocal photoluminescence scan of a section of the device, where the golden regions are part of the on-chip coplanar waveguide (CPW) used for applying quantum control pulses and NV centers appear as bright spots in between the conductors of the CPW. (**B**) Energy level diagrams of the NV center electron spin (left) and the electron spins in the bath (right). An applied magnetic field splits the NV spin triplet electronic ground state; the effective two-level system used here is formed by the spin sublevels $m_S = 0$ (labeled $|0\rangle$) and $m_S = -1$ (labeled $|1\rangle$) *(14)*. (**C**) Coherent driven oscillations of NV1. For the pulsed experiments the same Rabi frequency is used *(14)*. (**D**) Decay during free evolution of NV1 probed using Ramsey interference. Solid line is a fit *(14)*. The fast oscillating component is due to a detuning of the driving field of 15 MHz with respect to the spin transition, while the beating is caused by the hyperfine interaction with the host nuclear spin.

**Fig. 2. Optimized dynamical decoupling of NV1.**

(**A**) Left: state fidelities for CPMG decoupling sequence applied to NV1. The blue curve is a spin echo measurement. High state fidelity is recovered for increasing number of pulses *N*. Solid lines are fits to $\sim\exp[-(t/T_{\text{coh}})^3]$. Right: vertical lines indicate the location of π-pulses. (**B**) Comparison of decoupling with CPMG (orange) and UDD (green) for $N$ = 6 pulses. The solid lines are fits to $\sim\exp[-(t/T_{coh})^3]$. The right panel shows the $1/e$ decay times from fits to data and to simulations *(14)*. The same color scheme applies. (**C**) Single-axis decoupling for different input states, showing state-selective decoupling for the CPMG sequence with $N = 12$ operations (shown in the upper right). Bloch sphere on the right shows input states and the decoupling axis. Solid lines are numerical simulations incorporating the experimental pulse errors *(14)*. (**D**) Double-axis decoupling, with XY4 sequence with $N = 12$, showing excellent decoupling for both input states. Pulse timings are the same as for CPMG but with the decoupling axis alternating between X and Y, as is shown on the right. The simulations for $|x\rangle$ and $|y\rangle$ yield virtually the same curve and therefore appear as one.

**Fig 3. Universal decoupling demonstrated with Quantum Process Tomography.**
QPT is performed at free evolution times of 4.4, 10 and 24 μs for XY4 with $N = 8$ (see Fig. S2B). At $t = 4.4$ μs the measured process matrix nearly equals the identity process matrix $\chi$ (fidelity of $0.96 \pm 0.02$) indicating close-to-perfect quantum state protection. At longer free evolution times the process changes into pure dephasing in accordance with our model of the spin bath.

**Fig 4. Scaling of the coherence enhancement with number of control pulses.**
(**A**) Decoupling for different number of control pulses $N$. Increasing $N$ extends the coherence to longer times. Solid lines are simulations *(14)*. (**B**) Data rescaled to the normalized time axis $t/(T_2 N^{2/3})$. (**C**) Coherence $1/e$ decay time ($T_{coh}$) plotted as a function of the number of control pulses for NV1 and NV2. Solid lines are fits to $T_{coh}(N) = T_2\, N^{2/3}$ with $T_2$ as free parameter.

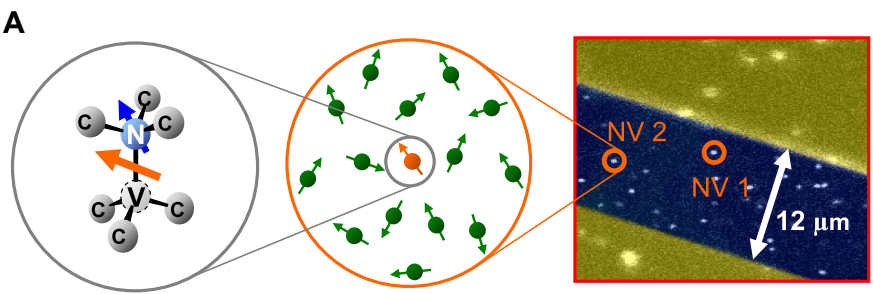

**A**

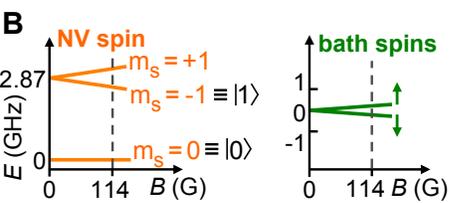

**B**

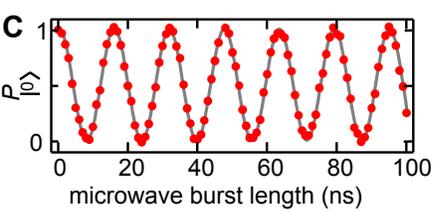

**C**

**D**

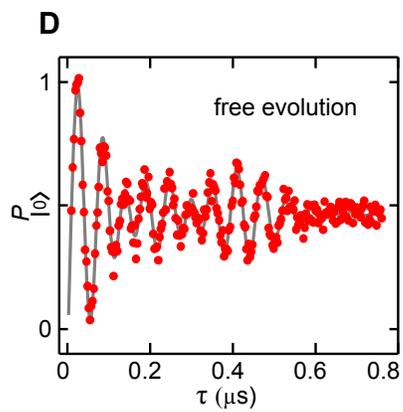

free evolution

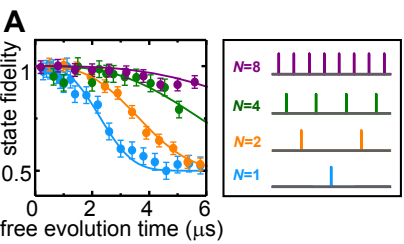
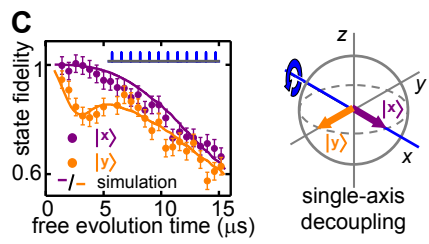
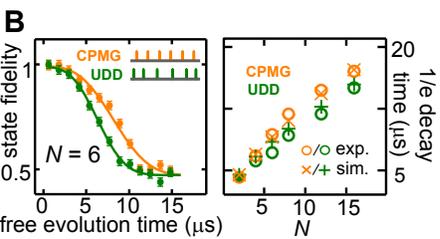
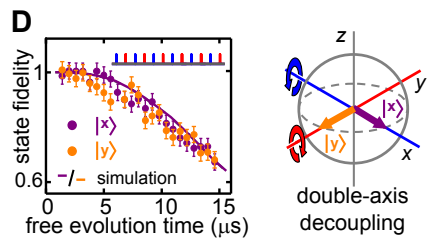

## Quantum Process Tomography

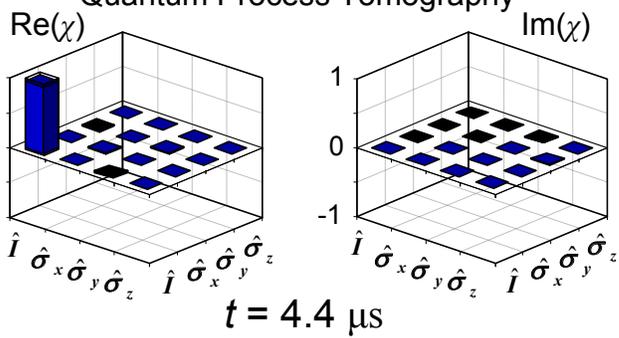

$t = 4.4$ μs

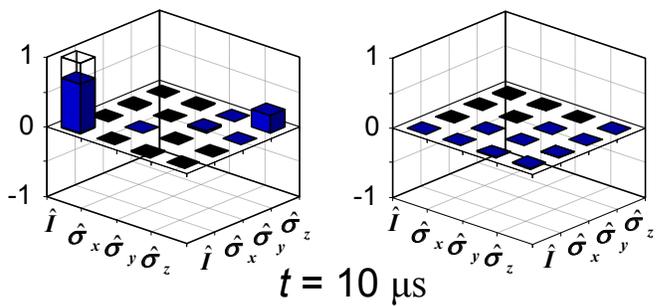

$t = 10$ μs

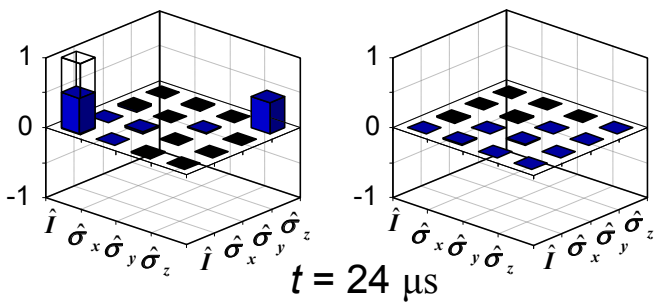

$t = 24$ μs

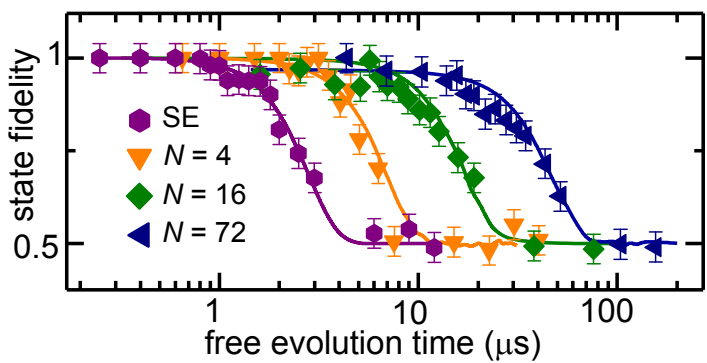
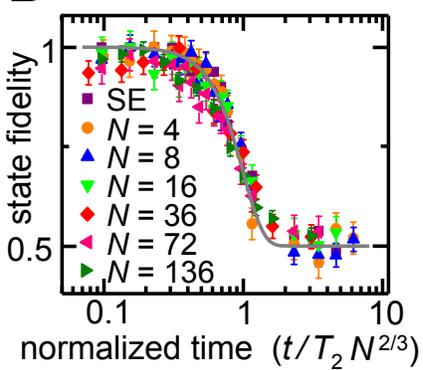
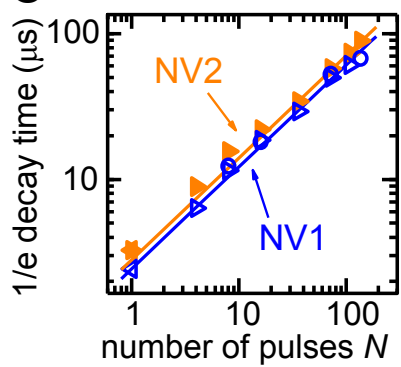